\documentclass[fleqn,twoside]{article}

\usepackage{cite}

\usepackage{espcrc2}
\usepackage{graphicx}

\readRCS
$Id: espcrc2.tex,v 1.2 2004/02/24 11:22:11 spepping Exp $
\usepackage{epsfig}

\newcommand{\AmS}{{\protect\the\textfont2
  A\kern-.1667em\lower.5ex\hbox{M}\kern-.125emS}}

\def\z#1{{\zeta_{#1}}}
\def\ca{{C_{\!A}}}
\def\cas{{C^{\, 2}_{\!A}}}
\def\cat{{C^{\, 3}_{\!A}}}
\def\cf{{C_F}}
\def\nf{{n^{}_{\! f}}}
\def\nfs{{n^{2}_{\! f}}}

\def\cfs{{C^{\:\! 2}_{\! F}}}
\def\cft{{C^{\:\! 3}_{\! F}}}

\def\caf{{C^{}_{\!A\!F}}}
\def\cafs{{C^{\,2}_{\!A\!F}}}
\def\caft{{C^{\,3}_{\!A\!F}}}

\def\flg11{fl^{\,g}_{11}}

\def\x1{{(1 \! - \! x)}}

\newcommand{\hspn}{{\hspace{-9mm}}}

\newcommand{\eqLL}{\raisebox{-0.07cm}{$\:\stackrel{{\rm LL}}{=}\:$} }

\newcommand{\beq}{\begin{equation}}
\newcommand{\eeq}{\end{equation}}
\newcommand{\bea}{\begin{eqnarray}}
\newcommand{\eea}{\end{eqnarray}}
\newcommand{\nn}{\nonumber}
\newcommand{\MSb}{$\overline{\mbox{MS}}$}
\newcommand{\as}{\alpha_{\rm s}}
\newcommand{\ar}{a_{\sf s}}
\newcommand{\ra}{\rightarrow}
\newcommand{\ep}{\varepsilon}

\title{
\vspace*{-18mm}
\rightline{
{\normalsize{LTH 880, DESY 10-121 (August 2010)}}}
\vspace*{+5mm}
On higher-order flavour-singlet splitting and coefficient functions at
large$\:x$%
{\thanks{Proceedings of the workshops~ {\it Loops and Legs in Quantum
Field Theory}, April 2010, W\"orlitz (Germany) and (shortened)
{\it DIS 2010}, Florence, April 2008.}}$\!\!\!\!\!$
}

\author{A. Vogt\address[UoL]{Department of Mathematical Sciences, University of
 Liverpool, Liverpool L69 3BX, United Kingdom},
 G. Soar\addressmark[UoL],
 S. Moch\address{Deutsches Elektronensynchrotron DESY, Platanenallee 6,
 D--15738 Zeuthen, Germany}
 and J.A.M. Vermaseren\address{NIKHEF, Science Park 105, 1098 XG Amsterdam,
 The Netherlands}}

\begin{document}

\begin{abstract}
We discuss the large-$x$ behaviour of the splitting functions $P_{\rm qg}$ and 
$P_{\rm gq}$ and of flavour-singlet coefficient functions, such as the gluon 
contributions $C_{2,\rm g}$ and $C_{L,\rm g}$ to the structure functions 
$F_{2,L}$, in massless perturbative~QCD. 
These quantities are suppressed by one or two powers of $\x1$ with respect to 
the $\x1^{-1}$ terms which are the subject of the well-known threshold 
exponentiation. 
We show that the double-logarithmic contributions to $P_{\rm qg}$, $P_{\rm gq}$ 
and $C_L$ at order $\as^{4}$ can be predicted from known third-order results 
and present, as a first step towards a full all-order generalization, the 
leading-logarithmic large-$x$ behaviour of $P_{\rm qg}$, $P_{\rm gq}$ and 
$C_{2,\rm g}$ at all orders in~$\as$.
\vspace*{-1mm}
\end{abstract}

\maketitle

\section{Introduction}

Inclusive deep-inelastic lepton-nucleon scattering (DIS) via the exchange of a 
colour-neutral (gauge) boson, for the basic kinematics see Fig.~1, is a 
benchmark process of perturbative QCD.
 
\begin{figure}[hb]
\vspace*{-4.5mm}
\parbox{3.7cm}{
\begin{center}
\vspace*{-1.4cm}
\includegraphics[bb = 170 560 430 360, scale = 0.4]{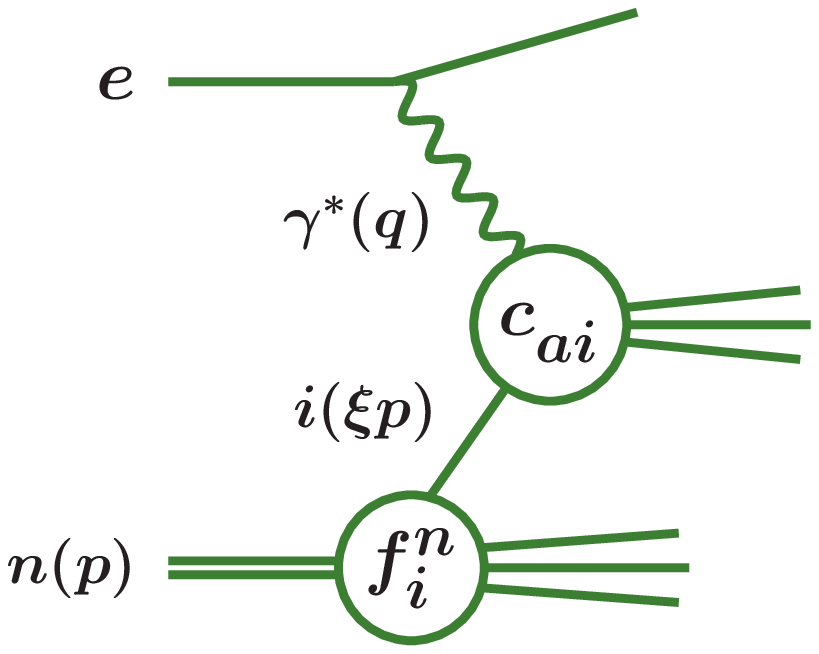}
\end{center}
}
\parbox{3.5cm}{
\vspace{1mm}
{Scale, Bjorken variable}
\begin{eqnarray*}
 ~~~Q^2\!\!  & = &  - q^2 \\[0.5mm]
 ~~~x        & = &  Q^2/(2\:\! p\cdot\! q)
\end{eqnarray*}
\vspace{1mm}
Lowest order$\,$: {$x = \xi$}}
\vspace*{-5mm}
\caption{\label{DISkin}
The kinematics and perturbative QCD factorization of photon-exchange DIS.}
\vspace*{-5mm}
\end{figure}

\noindent
Disregarding contributions suppressed by powers of $1/Q^2$, the structure 
functions in electromagnetic DIS are given by
\beq
\label{Fa-pQCD}
 ~x^{-1} F_a^{\:\!n}(x,Q^2) = 
 \big[ \, C_{a,i}(\as(Q^2)) \otimes f_{i}^{\,n} (Q^2) \big] (x)~~
\eeq
in terms of the coefficient functions $C_{a,i}$, $a = 2,\,L$, $i = \rm q, g$, 
and the nucleon parton distributions $f_{i}^{\,n\!}$. Here $\otimes$ denotes 
the standard Mellin convolution, and the summation over $i$ is understood. 
Without loss of information, we identify the renormalization and factorization 
scale with the physical scale $Q^2$ in Eq.~(\ref{Fa-pQCD}) and throughout this
article.

The scale dependence of the parton densities is
\beq
\label{qG-evol}
  ~\frac{d\:\! f_i^{}(\xi,\mu^2)}{d \ln Q^2} =
  \big[ \, P_{ik}(\as(Q^2)) \otimes f_k^{}(Q^2) \big] (\xi) \: .
\eeq
The coefficient functions in Eq.~(\ref{Fa-pQCD}) and the splitting functions
$P_{ik}$ can be expanded in powers of the strong coupling constant 
$\ar \equiv \as/(4\pi)$, 
\bea
\label{C-exp}
  ~C_{a,i}(x,\as) &\!\!=\!& {\textstyle \sum_{\,l=0}^{}} 
  \;\ar^{\,l + l_a}  c_{a,i}^{(l)}(x) \; ,
\\
  ~P_{ik}(x,\as)\,  &\!=\!& {\textstyle  \sum_{\,l=0}^{}} 
  \;\ar^{\,l+1} \, P_{ik}^{\,(l)}(x) 
\label{P-exp}
\eea
with $l_a = 0$ for $F_2$ (and the Higgs-exchange structure function $F_\phi$ 
discussed below), and $l_a = 1$ for the longitudinal structure function $F_L$.
In this notation, the N$^n$LO approximation includes the contributions with
$l \leq n$ in both Eqs.~(\ref{C-exp}) and~(\ref{P-exp}).

\vspace{1mm}
The above (spin-averaged) splitting functions are presently known to $n=2$ 
\cite{MVV3,MVV4}, i.e., the next-to-next-to-leading order (NNLO $\equiv$ 
N$^2$LO). The coefficient functions for the most important structure functions
(including $F_3$ for charge-averaged $W$-exchange) have also been 
fully computed to order $\as^3$ \cite{MVV5,MVV6,MVV10},
while the less important charge-asymmetry $W$-cases are available
only through a couple of low-integer Mellin-$N$ moments \cite{MR1,MRV1}. 
The frontier in the present massless case are now the $\as^{4}$ corrections,
for which first results have been obtained at the lowest value of $N$
\cite{BCh06,BChK10}.~See Ref.~\cite{BBK09} for the status of the third-order
computation of the heavy-quark contributions to DIS.

\section{The general large-$x$ behaviour} 

We are interested in the leading contributions, in terms of powers in $\x1$, 
to Eqs.~(\ref{C-exp}) and~(\ref{P-exp}). The form of the diagonal splitting
functions is stable under higher-order corrections in the \MSb\ scheme, 
viz~\cite{Korch89}
\beq
\label{P-diag}
  ~P_{ii}^{\,(l)} = A_{i}^{(l)} \x1_+^{-1}
    +  B_{i}^{(l)} \delta \x1 \, + \;\ldots \; .~
\eeq
The off-diagonal quantities, however, receive a double-logarithmic
higher-order enhancement,
\beq
\label{P-offd}
 ~P_{i \neq j}^{\,(l)}\, = {\textstyle  \sum_{a=0}^{2l}} \; 
 A_{ij,a}^{\,(l)}\ln^{\:\!2l-a} \!\x1 \: + \;\ldots \:\: ,
\eeq
where $A_{ij,a}^{\,(l)} \propto (\ca - \cf)^{\,l-a}$ for $a < l$ for (at least)
$l \leq 2$ \cite{MVV4}, i.e., all double logarithms vanish for $\cf = \ca$, 
which is part of the colour-factor choice leading to an ${\cal N}\! =\! 1$ 
supersymmetric theory.

\vspace{1mm}
The leading large-$x$ parts of `diagonal' coefficient functions, e.g.,
$C_{2,\rm q}$ and $C_{\phi,\rm g}$, are given by
\beq
\label{C-diag}
 ~c_{\rm diag}^{\,(l)}\, = {\sum_{a=0}^{2l-1}} \;
 D_{i,a}^{\,(l)} \bigg[ {\ln^{\:\!2l-1-a} \!\x1 \over 1\!-\! x} \bigg]^{-1}_+
 \: + \;\ldots \;.~~
\eeq
These terms are resummed by the soft-gluon exponentiation 
\cite{SoftGlue}. For DIS structure functions 
(and some other semi-leptonic processes) this resummation is known at 
the next-to-next-to-next-to-leading logarithmic accuracy, i.e., the highest 
six logs are completely known to all orders \cite{MVV7}.

\vspace{1mm}
No resummation has been derived so far for the off-diagonal (flavour-singlet) 
coefficient functions such as $C_{2,\rm g}$ and $C_{\phi,\rm q}$ which are of 
the form
\beq
\label{C-offd}
 ~c_{\rm off-d}^{\,(l)}\, = {\textstyle \sum_{a=0}^{2l-1}} \;
 O_{i,a}^{\,(l)} \ln^{\:\!2l-1-a} \!\x1 \: + \;\ldots \;.~~
\eeq
The coefficient functions for $F_L$ are suppressed by one power in $\x1$ 
with respect to those of $F_2$,

\vspace{-1.5mm}
\beq
\label{C-FL}
 ~c_{L,i}^{\,(l)}\, = {\sum_{a=0}^{2l}} \;
 L_{L,i}^{\,(l)} \x1^{\delta_{ig}} \ln^{\:\!2l-a} \!\x1 \: + \;\ldots \;,~
\eeq
recall our notation with $l_L=1$ in Eq.~(\ref{C-exp}). The double-log 
contributions to $C_{L,\rm q}$ (and the $\cf = 0$ part of $C_{L,\rm g}$) have 
been resummed in Ref.~\cite{MV3}, i.e., the respective highest three logarithms 
($a = 0,\,1$ and 2 in Eq.~(\ref{C-FL})) are known to all orders.

\vspace{1mm}
Our aim is to derive corresponding predictions for all quantities in 
Eqs.~(\ref{P-offd}), (\ref{C-offd}) and (\ref{C-FL}). The present contribution 
is a brief status report of this programme, which has not been finished so far.

\section{Physical evolution kernel for \boldmath ($F_2, F_\phi$)} 

The results of Ref.~\cite{MV3} and their \mbox{extension} to the non-leading 
corrections for $C_{2,\rm q}$ and other quantities at all orders in $\x1$
\cite{MV5}, see Ref.~\cite{smvvRC09} for a brief summary, have been obtained by 
studying the non-singlet physical evolution kernels for the respective 
observables. It is thus natural to study also flavour-singlet physical kernels.

\vspace{1mm}
The most natural complement to the standard quantity $F_2$ with $c_{2,i}^{(0)} 
= \delta_{iq}\, \delta\x1$ is a structure function for a probe which directly
interacts only with gluons, such as a scalar $\phi$ with a $\phi\, G^{\mu\nu}
G_{\mu\nu}$ coupling to the gluon field \cite{FP82}. In the Standard Model
this interaction is realized for the Higgs boson in the limit of a heavy
top-quark \cite{HGGeff1,HGGeff2}. The coefficient functions $C_{\phi,i}$ have
been determined recently in Refs.~\cite{DGGL} and \cite{SMVV1} to the second 
and third order in $\as$, respectively.

\vspace{1mm}
We thus consider the 2-vector singlet structure function and $2\!\times\!2$ 
coefficient-function matrix
\beq
\label{F2Fphi}
  ~F \:=\:
  \Big( \begin{array}{c} \!\! F_2 \!\! \\ \! F_\phi \!\!
        \end{array} \Big)
\;\; , \;\;\;
  C \:=\:
  \Big( \! \begin{array}{cc} C_{2,\rm q}^{}\! & C_{2,\rm g}^{}\! \\
  C_{\phi,\rm q}\! & C_{\phi,\rm g}\! \end{array}\! \Big) 
  \;\; .
\eeq
With $P$ denoting the matrix of the splitting functions (\ref{C-diag}) and
(\ref{C-offd}), the evolution kernel for $F$ reads
\bea
\label{F-evol}
  ~\frac{d F}{d \ln Q^2} &\!\!\! =\!\!\! &
  \frac{d\, C}{d \ln Q^2}\, f \, + \, C P f 
\nn \\[0.5mm] &\!\!\! =\!\!\! &
  \Big( \beta(\ar)\, \frac{d\, C}{d \ar}\,C^{\,-1}  + C P C^{\,-1} \Big) F
\\ &\!\!\! =\!\!\! &
  K F \;\; \mbox{ with } \;\;
 K \:=\:
 \Big( \! \begin{array}{cc} K_{22}\! & K_{2\phi}\! \\
  K_{\phi 2}\! & K_{\phi\phi}\! \end{array}\! \Big)
  \;\: . \nn
\eea
$\beta(\ar) = - \beta_0 \ar^2 + \ldots$ with $\beta_0 = 11\,\ca/3 - 2\,\nf/3$~%
is the standard beta function of QCD. All products of $x$-dependent quantities 
have to be read as convolutions (or products of their Mellin transforms).

\vspace{1mm}
After expanding in $\as$, the first term in the second line of 
Eq.~(\ref{F-evol}) receives double-logarithmic contributions from the 
non-singlet and singlet coefficient functions (\ref{C-diag}) and (\ref{C-offd}).
The second term, absent in the non-singlet cases of Refs.~\cite{MV3,MV5}, 
includes also the double-log terms of Eq.~(\ref{P-offd}). 

\vspace{1mm}
The crucial observation, proven by available three-loop calculations to order 
$\as^4$ for the non-singlet parts (thanks to Eq.~(\ref{P-diag})) and to order
$\as^3$ for the singlet contribution, is that the physical kernel $K$ is only 
single-log enhanced~\cite{SMVV1}, i.e., 

\vspace{-2mm}
\beq
\label{KabLL}
 ~K_{ab}^{(l)} = {\sum_{\eta=0}^{l}} \:
 A_{ab,\eta}^{\,(l)} \x1^{-\delta_{ab}} \ln^{\:\!l-\eta} \!\x1 \: + \;\ldots~ 
\eeq
 
\vspace{-2mm}
\noindent
where the expansion coefficients $K_{ab}^{(l)}$ are defined as in 
Eq.~(\ref{P-exp}) for the splitting functions above.

\vspace{0.5mm}
We conjecture that also the flavour singlet part remains single-log enhanced at
the fourth order. This implies a cancellation between the double-logarithmic
contributions to the, so far unknown, off-diagonal $l=3$ splitting functions 
(\ref{P-offd}) and the known \cite{MVV6,SMVV1} coefficient functions 
to order $\as^3$ from which the former can be deduced. The results are 
\bea
\label{Pqg3L}
 P^{\,(3)}_{\rm qg}/\nf & \!\! = \!\! &
      \ln^6 \! \x1 \* \: \cdot 0
\nn \\ & & \mbox{\hspn\hspn}
     + \ln^5 \! \x1 \* \Big [ \,
            {22 \over 27}\, \* \caft 
          - {14 \over 27}\, \* \cafs \* \cf 
          + {4 \over 27}\, \* \cafs \* \nf
          \Big]
\nn \\[0.4mm] & & \mbox{\hspn\hspn}
     + \ln^4 \! \x1 \* \Big [ \,
            \Big( {293 \over 27}
            - {80 \over 9}\, \* \z2\! \Big) \* \caft 
          - {116 \over 81}\, \* \cafs \* \nf
\nn \\[0.4mm] & & \mbox{\hspn}
          + \Big( {4477 \over 16}
          - 8 \* \z2\! \Big) \* \cafs \* \cf 
          - {13 \over 81}\, \* \caf \* \cfs 
\nn \\[0.4mm] & & \mbox{\hspn}
          + {17 \over 81}\, \* \caf \* \cf \* \nf
          - {4 \over 81}\, \* \caf \* \nfs
          \Big]
\nn \\[0.4mm] & & \mbox{\hspn\hspn}
      + {\cal O} \left( \ln^3 \! \x1 \right) 
\;\; ,
\\[1mm]
\label{Pgq3L}
 P^{\,(3)}_{\rm gq}/\cf & \!\! = \!\! &
      \ln^6 \! \x1 \* \: \cdot 0
\nn \\ & & \mbox{\hspn\hspn}
     + \ln^5 \! \x1 \* \Big [ \,
            {70 \over 27}\, \* \caft 
          - {14 \over 27}\, \* \cafs \* \cf
          - {4 \over 27}\, \* \cafs \* \nf
          \Big]
\nn \\[0.4mm] & & \mbox{\hspn\hspn}
     + \ln^4 \! \x1 \* \Big [ \,
            \Big( {3280 \over 81}
            + {16 \over 9}\, \* \z2\! \Big)\, \* \caft 
          - {256 \over 27}\, \* \cafs \* \nf
\nn \\[0.4mm] & & \mbox{\hspn}
          + \Big( {637 \over 18}
          - 8 \* \z2\! \Big)\, \* \cafs \* \cf
          - {49 \over 81}\, \* \caf \* \cfs
\nn \\[0.4mm] & & \mbox{\hspn}
          + {17 \over 81}\, \* \caf \* \cf \* \nf
          + {32 \over 81}\, \* \caf \* \nfs
          \Big]
\nn \\[0.4mm] & & \mbox{\hspn\hspn}
      + {\cal O} \left( \ln^3 \! \x1 \right)
\eea
with $\caf \equiv \ca - \cf$. 
The vanishing of the leading $\ln^6 \x1$ contributions is due to a cancellation
of contributions. Below we will address the question whether this cancellation 
is accidental or a structural feature.
Eqs.~(\ref{Pqg3L}) and (\ref{Pgq3L}) show the colour-factor pattern already 
noted for $l \leq 2$ below Eq.~(\ref{P-offd}). The feature is not an obvious 
consequence of our derivation and can thus be viewed as a non-trivial check of
the above conjecture.

\vspace{0.5mm}
The extension of the above results to all powers of $\x1$
can be found in Ref.~\cite{SMVV1}.

\section{Physical evolution kernel for \boldmath ($F_2, F_L$)}

The system of standard DIS structure functions$\!\!$
\beq
\label{F2FL}
  ~F =
  \Big( \begin{array}{c} \!\! F_2 \!\! \\ \! \widehat{F}_L \!\!
        \end{array} \Big)
\;\; , \;\;\;
  \widehat{F}_L = F_L / \big(\ar\, c_{L,\rm q\,}^{\,(0)} \big) \;\; ,
\eeq
studied before in Refs.~\cite{Catani96,BRvN00}, can be analyzed in complete 
analogy to the previous section. Our normalization of $\widehat{F}_L$ (of 
course Eq.~(\ref{F2FL}) involves a simple division only in Mellin-$N$ space) 
leads to
\beq
\label{CL-hat}
 ~C =
 \bigg( \!\begin{array}{cc} \!\delta \x1 \!\!& 0 \\
                            \!\delta \x1 \!\!& \hat{c}_{L,\rm g}^{\,(0)}\!
  \end{array}\! \bigg)
  + \: \sum_{l=1} \ar^l \:
  \Bigg( \! \begin{array}{cc} \!c_{2,\rm q}^{(l)}\! &\!\! c_{2,\rm g}^{(l)}\!
  \\[1mm]\!\hat{c}_{L,\rm q}^{\,(l)} &\!\! \hat{c}_{L,\rm g}^{\,(l)}\!
  \end{array}\! \Bigg) \, .\:
\eeq
The resulting elements of the physical kernel
\beq
\label{K-F2FL}
 ~K =
 \bigg( \! \begin{array}{cc} \! K_{22}^{}\!\! & \!\! K_{\rm 2L}^{}\!\!\\[0.5mm]
  \! K_{\rm L2}^{}\!\! & \!\! K_{\rm LL}^{}\!\! \end{array}\! \bigg)
\eeq
are again single-log enhanced at large $x$ and read

\vspace{-2mm}
\beq
 ~K_{ab}^{(l)} =
{\sum_{\eta=0}^{l}} \;
 \widehat{A}_{ab,\eta}^{\,(l)} \x1^{-1} \ln^{\:\!l-\eta} \!\x1 \: + \;\ldots~
\eeq
 
\vspace{-2mm}
\noindent
at, at least, $l \leq 3$ for the upper row of Eq.~(\ref{K-F2FL}), with  
$\widehat{A}_{2\rm L,0}^{\,(l)} = 0$, and at $l\leq 2$ for the lower row. 

\vspace{1mm}
Conjecturing that this behaviour holds at $l = 3$ also for
$K_{L2}$ and $K_{LL}$, the three-loop results of 
Refs.~\cite{MVV3,MVV4,MVV5,MVV6} together with Eq.~(\ref{Pqg3L}) yield
\bea
\label{cLq3}
  ~c^{\,(3)}_{L,\rm q}\, / \cf\!  & \!\! = \!\! &
      \ln^6 \! \x1 \* \; {16 \over 3}\: \cft
\nn \\[-0.4mm] & & \mbox{\hspn\hspn}\;
    + \ln^5 \! \x1 \* \Big [ \,
    ( 72 - 64\, \z2 ) \,\cft
    + {80 \over 9}\, \cfs \nf
\nn \\ & & \mbox{}\;
    - \Big( \, {728 \over 9} - 32\, \z2 \Big) \: \cfs \ca \Big]
\nn \\[0.4mm] & & \mbox{\hspn\hspn}\;
    + \ln^4 \! \x1 \* \cdot \big[ \, \mbox{known coefficients} \,\big]
\nn \\[0.4mm] & & \mbox{\hspn\hspn}\; 
    + {\cal O} \left( \ln^3 \! \x1 \right)
\;\; ,
\\[1mm]
\label{cLg3}
  ~c^{\,(3)}_{L,\rm g}\, / \nf\!  & \!\! = \!\! &
      \x1 \ln^6 \! \x1 \* \; {32 \over 3}\: \cat
\nn \\ & & \mbox{\hspn\hspn}\;
    + \x1 \ln^5 \! \x1 \* \Big [ 
          - {2080 \over 9}\, \* \cat 
          + {64 \over 9}\, \* \cas \* \nf 
\nn \\[0.8mm] & & \mbox{\hspn\hspn}\;\;\:
          + {104 \over 3}\, \* \cas \* \cf
          + {40 \over 3}\, \* \cft \Big]
\nn \\[0.8mm] & & \mbox{\hspn\hspn}\;
    + \x1 \ln^4 \! \x1 \* 
\Big[
         \Big( \, {70760 \over 27} - 352\,\* \z2 \Big) \* \cat
\nn \\[0.8mm] & & \mbox{\hspn\hspn} \;\;\:
         - \Big( {25306 \over 27} - {320 \over 3}\,\* \z2 \Big) \* \cas \* \cf
         - {4192 \over 27}\: \* \cas \* \nf
\nn \\ & & \mbox{\hspn\hspn} \;\;\:
         + \Big( {1600 \over 27} + 32\,\z2 \Big) \* \ca \* \cfs
         + {556 \over 27}\: \* \ca \* \cf \* \nf
\nn \\[0.8mm] & & \mbox{\hspn\hspn} \;\;\:
         + {32 \over 27}\: \* \ca \* \nfs
         + \Big( 38 - {320 \over 3}\,\* \z2 \Big) \* \cft
         + {308 \over 27}\: \* \cfs \* \nf
\Big]
\nn \\[0.4mm] & & \mbox{\hspn\hspn}\;
    + {\cal O} \left( \x1 \ln^3 \! \x1 \right)
\; ,
\eea
where the coefficient of $\ln^4 \! \x1$ in Eq.~(\ref{cLq3}) has been suppressed
for brevity. The complete form of this equation has been given in 
Ref.~\cite{MV3}, where it was derived in another manner which did not involve 
the off-diagonal splitting functions. Consequently the consistency of the two
derivations provides another confirmation of the correctness of the above
result for $P^{\,(3)}_{\rm qg}$. The non-$\cf$ parts of Eq.\ (\ref{cLg3}) --
here, as App.~C of Ref.~\cite{SMVV1}, given for $W$-exchange, i.e., without 
the $\flg11$ contribution for the photon case \cite{MVV6} --
have also been derived, but not explicitly written down, in Ref.~\cite{MV3}.

\vspace*{-0.5mm}
\section{Unfactorized off-diagonal amplitudes}

The single-log enhancement of the above physical kernels suggests an iterative 
structure of the unfactorized structure functions (forward  amplitudes), from 
which the splitting and coefficient functions are obtained in Mellin-$N$ space 
via the mass-factorization relations
\beq
\label{TCZgam}
  ~T_{a, j} \,=\, \widetilde{C}_{a, i} \, Z_{\,ij}
\: , \;\;
  - \:\!\gamma \equiv P =
 \frac{d\:\! Z }{d\ln Q^{\,2} } \, Z^{-1}
\:\: .
\eeq
The $D$-dimensional coefficient functions $\widetilde{C}_a$ include terms with 
$\ep^k$, $k \geq 0$ in dimensional regularization with $D = 4 - 2\ep$. 
The transition functions $Z$ collect all terms which are singular for~%
$\ep \ra 0$.  Inverting the second relation in Eq.~(\ref{TCZgam}) yields
\bea
\label{Zn-exp}
  ~Z^{} \!\left|_{\ar^n} \right. \!\!
  &\!\!\!=\!\!\!& {1 \over \ep^n} {\gamma_{\:\!0}^{\,n}  \over n!}
  + \ldots
\\[1mm] & & \mbox{\hspn} \!\!
  + {1 \over \ep^2}
    \bigg( { \gamma_{\:\!0}^{} \gamma_{\:\!n-2}^{} \over n (n-1)}
    + { \gamma_{\:\!n-2}^{} \gamma_{\:\!0}^{} \over n } + \ldots \bigg)
  + {1 \over \ep} { \gamma_{\:\!n-1}^{} \over n}
\; .  \nn
\eea
At order $\as^n$, the $\ep^{-n} \ldots \ep^{-2}$ contribution to $T_a$ are 
given in terms of lower-order terms. The~$\ep^{-n}$ and $\ep^{0}$ coefficients
include the $n$-loop 
splitting functions and (four-dimensional) coefficient functions $C_a$, 
respectively. Terms with $\ep^{\:\!k\!},\: 0 < k < l\:\!$ are required for 
the factorization at order $n\!+\!l$.

\vspace*{1mm}
We now focus on the leading-logarithmic (LL) contributions to the off-diagonal
quantities $T_{\phi, \rm q}$ and $T_{2, \rm g}$ and summarize the results of
Ref.~\cite{AV2010}.

\vspace{1mm}
\noindent
With $\,L \equiv \ln N\:\!$ these terms are of the form
\bea
\label{T-offd}
    ~T_{\phi, \rm q}^{\,(n)}/ \cf\!
  &\!\!\! \eqLL \!\!\!& 
    T_{2, \rm g}^{\,(n)}/ \nf
  \;\eqLL\;  
  {L^{n-1} \over N \ep^n} \sum_{k=0}^\infty\, (\ep L)^k {\cal L}_{n,k} \cdot
\nn \\ & & \mbox{\hspn} \cdot\:
  \left( C_{\!F}^{\,n\!-\!1\!} + C_{\!F}^{\,n\!-\!2} \ca + \ldots
       + C_{\!A}^{\,n\!-\!1} \right)
\; ,
\eea
i.e., the coefficients ${\cal L}_{n,k}$ are the same for both amplitudes and
all combinations of $\cf$ and $\ca$. Consequently an all-order relation for one
colour structure of either amplitude is sufficient. 
The calculations of Ref.~\cite{MVV4,SMVV1} imply such a relation, 
\beq
\label{Tphiq}
  ~T_{\phi, \rm q}\big|_{C_{\!F}\: \rm only} \;\eqLL\:\:
  T_{\phi, \rm q}^{\,(1)} \;
  { \exp (\ar T_{2, \rm q}^{\,(1)}) - 1\over T_{2, \rm q}^{\,(1)} }
  \:\: ,
\eeq
in terms of the first-order expressions known to all powers of $\ep$,
\bea
\label{Tphiq1}
 ~T_{\phi, \rm q}^{\,(1)} &\!\!\!\eqLL\!\!\!&
  - {2\:\! \cf \over N} \,{1 \over \ep}\, \exp (\ep \ln N)
\:\: , \\
\label{T2g1}
  ~T_{2, \rm q}^{\,(1)} &\!\!\!\eqLL\!\!\!&
  4\:\! \cf \,{1 \over \ep^2}\, ( \exp (\ep \ln N) - 1 )
\:\: .
\eea

After carrying out the mass factorization to a very high order in $\as$ 
(using {\sc Form} \cite{FORM}), the all-order analytic expressions for the 
leading-logarithmic contributions to the splitting functions and coefficient 
functions have been derived. The former quantities are given by
\beq
\label{PqgLL}
  ~P_{\rm qg}^{\:\rm LL}(N,\as) \;=\;
  {\nf \over N}\: {\as \over 2\pi}\; {\cal B}_{0} (\tilde{a}_{\rm s} )
\eeq
with
\bea
\label{B0}
  ~{\cal B}_{0}(x) &\!\!\!=\!\!\!&
   \sum_{n=0}^\infty \,\frac{B_n}{(n!)^2} \: x^{n}
\nn \\ &\!\!\!=\!\!\!&
  1 \,-\, {x \over 2}
  \: - \:\sum_{n=1}^\infty \,\frac{(-1)^n}{(2n)!^2} \; |B_{2n}|\, x^{2n}
  \; ,
\eea
\beq
\label{atilde}
  ~\tilde{a}_{\rm s} \:=\: \as/\pi \; (\ca\! - \:\!\!\cf)  \ln^2 \!N
 \; .
\eeq
The coefficients $B_n$ in Eq.~(\ref{B0}) are the Bernoulli numbers in the 
standard normalization of Ref.\ \cite{AbrSteg}. The corresponding result for
$P_{\rm gq}$ is obtained by replacing $\nf$ by $\cf$ in Eq.~(\ref{PqgLL}),
and exchanging $\ca$ and $\cf$ in Eq.~(\ref{atilde}).

\vspace{1mm}
Due to $B_{2n+1} =\, 0$ for $n \geq 1$, the LL coefficients vanish at all even
orders in $\as$. Consequently the first lines of Eqs.~(\ref{Pqg3L}) and
(\ref{Pgq3L}) (recall the power counting (\ref{P-exp})) are not at all 
accidental.

\vspace{1mm}
The corresponding all-order result for the coefficient function $C_{2,\rm g}$ 
reads
\bea
\label{C2gLL}
  ~C_{2,\rm g}^{\:\rm LL}(N,\as) &\!\!\! = \!\!\!& 
  {1 \over 2N \ln N} \:{\nf \over \ca - \cf} \: \cdot
\\[0.5mm] & & \mbox{\hspn\hspn\hspn} \cdot
  \left\{ \exp \,(2\,\cf \ar \ln^2\! N) \, {\cal B}_{0} (\tilde{a}_{\rm s})
  - \exp \,(2\, \ca \ar \ln^2\! N) \right\}
\nn
\eea
$C_{\phi,q}^{\,\rm LL}$ can be obtained from this result by a simple
substitution of colour factors. 

\vspace*{1mm}
Inserting Eqs.~(\ref{P-diag}) (only the lowest-order terms with 
$A_q^{(0)} = 4\,\cf$ and $A_g^{(0)} = 4\,\ca$ contribute), (\ref{PqgLL}), 
(\ref{C2gLL}), and the well-known relation \cite{C2qLL}
\beq
\label{C2res}
 ~C_{2,\rm q} \;\eqLL \;\exp \left( 2 \ar\, \cf \ln^2 N \right)
\eeq
and its analogue for $C_{\phi,\rm g}$ into the physical kernel $K_{2\phi}$ in 
Eq.~(\ref{F-evol}), one finds that the highest double logarithm indeed vanishes
at all orders in $\as$. The same is found for $K_{\phi 2}$. Hence the
amplitude-based resummation verifies the conjecture made below 
Eq.~(\ref{KabLL}) for $a \neq b$, if presently only for the leading double 
logarithm.

\vspace*{1mm}
The function ${\cal B}_{0}(x)$ in Eq.~(\ref{B0}) appears to be a new 
function. The relation between $|B_{2n}|$ in the second line and the even
values of Riemann's $\zeta$-function \cite{AbrSteg} implies that this series
converges for all values of $x$. At positive $x$ the even part of 
${\cal B}_{0}$ compensates the odd $-x/2$ contribution up to an oscillation 
around zero, which persists, in an increasingly irregular manner, at very large
(and possibly all) values of $x$ \cite{DBpc}.

\begin{figure}[bth]
\vspace*{8mm}
\hspace*{2.3cm}{\large ${\cal B}_{0}(x)$}
\vspace*{-2.0cm}

\centerline{\epsfig{file=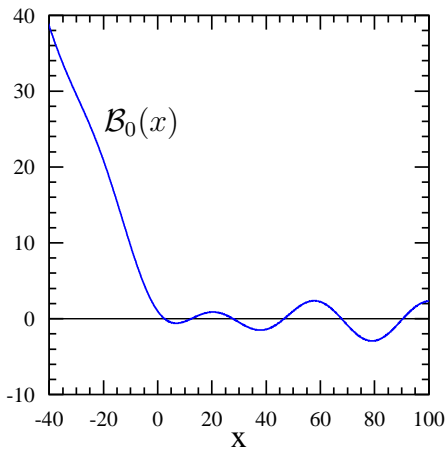,width=6cm,angle=0}}
\vspace{-10mm}
\caption{ 
The function ${\cal B}_{0}(x)$ in Eq.~(\ref{B0}), evaluated using its 
defining Taylor expansion.}
\vspace*{-10mm}
\end{figure}

\section{Summary and Outlook}

We have summarized the status of our \mbox{large-$x$} predictions of 
higher-order off-diagonal splitting functions and DIS coefficient functions. 
The coefficients of the highest three powers of $\ln \x1$ have been derived 
for the four-loop contributions to the splitting functions $P_{\rm qg}$ and 
$P_{\rm gq}$ from the three-loop coefficient functions and the 
single-logarithmic enhancement of the physical evolution kernel for the system 
($F_2,\, F_\phi$) of flavour-singlet structure functions at order $\as^4$ 
\cite{SMVV1}. 
In the present contribution we have employed these results to derive also the 
leading three large-$x$ logarithms for the fourth-order gluon coefficient 
function $C_{L,\rm g}$ for the longitudinal structure function from the 
analogous kernel for ($F_2,\, F_L$).

\vspace{0.5mm}
These results will become phenomenologically relevant, via effective $x$-space 
parametrizations analogous to, e.g., those of Ref.~\cite{NV4}, once the next 
major step towards a full fourth-order calculation of deep-inelastic 
scattering, the extension of Ref.~\cite{Mom3loop1} to order $\as^4$, has been 
taken.

\vspace{0.5mm}
The determination of flavour-singlet quantities from the physical kernels is
neither rigorous, nor -- unlike in flavour non-singlet cases \cite{MV3,MV5} -- 
can it be extended to all orders in $\as$. We
have presented first all-order leading-logarithmic results of a rigorous and 
more powerful approach, the prediction of the coefficients of the highest 
double logarithms from the $D$-dimensional structure of the unfactorized 
structure functions together with mass-factorization to all orders 
\cite{AV2010}.

\vspace{0.5mm}
We expect that, similar to the non-singlet case, the all-order resummation of
double-logarithmic large-$x$ contributions to flavour-singlet quantities can be
extended beyond parton evolution and inclusive DIS.
For example, the leading-logarithmic results of Ref.~\cite{AV2010} can be 
carried over directly to semi-inclusive electron-positron annihilation and 
$Z$- or Higgs-boson decay.
On the other hand, we do not foresee an extension of our results to 
single-logarithmically enhanced large-$x$ terms. It will be interesting to see 
whether such an extension can be achieved in alternative approaches such as the
application of soft-collinear effective theory to large-$x$ DIS 
\cite{SCET} or the recent path-integral formulation of Ref.~\cite{LSW08}.

\end{document}